# Watching the Generative AI Hype Bubble Deflate

*While the Generative AI hype bubble is slowly deflating, its harmful effects will last.*

**David Gray Widder**
Digital Life Initiative, Cornell University, New York City

**Mar Hicks**
School of Data Science, University of Virginia, Charlottesville

Only a few short months ago, Generative AI was sold to us as inevitable by the leadership of AI companies, those who partnered with them, and venture capitalists. As certain elements of the media promoted and amplified these claims, public discourse online buzzed with what each new beta release could be made to do with a few simple prompts. As AI became a viral sensation, every business tried to become an AI business. Some businesses added "AI" to their names to juice their stock prices,[1] and companies talking about "AI" on their earnings calls saw similar increases.[2]

Investors and consultants told businesses not to get left behind. Morgan Stanley positioned AI as key to a $6 trillion opportunity.[3] McKinsey hailed Generative AI as "the next productivity frontier" and estimated 2.6 to 4.4 trillion dollar gains,[4] comparable to the entire yearly GDP of the United Kingdom,[5] or all the world's agricultural production.[6] McKinsey also conveniently offers consulting services to help businesses "create unimagined opportunities in a constantly changing world."[7] Readers of this piece can likely recall the times they've been exhorted by news media or even leadership in their own industries and institutions to "learn AI," along with targeted ads hawking AI "boot camps."

While some have been wise to the hype for some time,[8] global financial institutions[9] and venture capitalists[10] are now also beginning to ask if Generative AI is over-hyped. In this essay, we argue that even as the Generative AI hype bubble slowly deflates, its harmful effects will last: carbon can't be put back in the ground, workers continue to need to fend off AI's disciplining effects, and the poisonous effect on our information commons will be hard to undo.

---

[1] Benzinga, "Stocks With 'AI' In the Name Are Soaring: Could It Be The Next Crypto-, Cannabis-Style Stock Naming Craze?"
[2] Wiltermuth, "AI Talk Is Surging during Company Earnings Calls — and so Are Those Companies' Shares."
[3] Morgan Stanley, "The $6 Trillion Opportunity in AI."
[4] Chui, Michael et al., "The Economic Potential of Generative AI: The next Productivity Frontier."
[5] $3.34 T in 2023, per World Bank, "United Kingdom."
[6] Estimated $4.59 T in 2024, per Statista, "Agriculture - Worldwide."
[7] McKinsey & Company, "Quantum Black - AI by McKinsey."
[8] See, for example: Bender and Hanna, "Mystery AI Hype Theater 3000"; Marcus, "The Great AI Retrenchment Has Begun"; Marx, "The ChatGPT Revolution Is Another Tech Fantasy"; Hanna, "The Grimy Residue of the AI Bubble."
[9] See, for example: Nathan, Grimberg, and Rhodes, "Gen AI: Too Much Spend, Too Little Benefit?"
[10] See, for example: Cahn, "AI's $600B Question."

## Historical hype cycles in the digital economy

Attempts to present AI as desirable, inevitable, and as a more stable concept than it actually is[11] follows well-worn historical patterns: one of the most important ways for a technology to gain market share and buy-in is to present it as an inevitable and necessary part of future infrastructure, and in turn to encourage the adaptation or building of new, anticipatory infrastructures around it. From the early history of the automobile and railroads, to the history of electricity and computers, we see the importance of this dynamic. All of these technologies required major infrastructure investments in order to become functional and dominant (roads, tracks, electrical grids, office labor and workflow changes) and none were inevitable, even though they may appear so in retrospect.[12]

The well-known phrase "nobody ever got fired for buying IBM" is a good, if partial, historical analogue to the current feeding frenzy to buy into AI: IBM, while expensive, was a recognized leader in automating workplaces, ostensibly to the advantage of those corporations. IBM famously re-engineered the environments into which its systems were installed, ensuring that office infrastructures and workflows were optimally reconfigured to mold to their computers, rather than the other way around. Similarly, AI corporations have repeatedly claimed that we are in a new age of not just *adoption* but of necessarily proactive *adaptation* to their new technology. Ironically, in AI waves past, IBM themselves have over-promised and under-delivered: some described their "Watson AI" product as a "mismatch" for the healthcare context they had sold it for, while others described it as "dangerous."[13] Time and again, AI has been crowned as an inevitable "advance" despite its many problems and shortcomings: from built-in biases, to inaccurate results, to privacy and intellectual property violations, to voracious energy use.

Nevertheless, in the media and–early on at least–among investors and corporations seeking to profit from it, AI has been publicly presented as unstoppable.[14] This was a key form of rhetoric coming from people eager to pave the way for a new set of heavily-funded technologies, not a statement of fact about the technology's robustness, utility, or even its likely future utility. Rather, it was a standard stage that we see in the development of many technologies. At this stage, a technology's manufacturers, boosters, and investors attempt to make it indispensable by integrating it, often prematurely, into existing infrastructures and workflows, counting on this entanglement to "save a spot" for the technology to be more fully integrated in the future. The more far-reaching this early integration, the more difficult it will be to disentangle or roll back the attendant changes–meaning that even broken or substandard technologies stand a better chance of becoming entrenched.[15]

---

[11] Suchman, "The Uncontroversial 'Thingness' of AI."
[12] See, for example: Oldenziel, Sousa, and van Wesemael, "Designing (Un)Sustainable Urban Mobility from Transnational Settings, 1850-Present" for cars; Nye, *Electrifying America* for electricity; Hicks, *Programmed Inequality* for computers; Burrell, "Artificial Intelligence and the Ever-Receding Horizon of the Future" for AI.
[13] Strickland, "How IBM Watson Overpromised and Underdelivered on AI Health Care."
[14] See, for example: Taylor, "Rise of Artificial Intelligence Is Inevitable but Should Not Be Feared, 'Father of AI' Says"; Shapiro, "Artificial Intelligence"; Raasch, "In Education, 'AI Is Inevitable,' and Students Who Don't Use It Will 'Be at a Disadvantage': AI Founder."

In the case of AI, however, as in the case of many other recent technology booms or boomlets (from the blockchain[16] to the metaverse[17] to clunky VR goggles[18]), this stage was also accompanied by severe criticism of both the rhetorical positioning of the technology as indispensable and of the technology's current and potential states. Historically, this form of critique is an important stage of technological development, and one in which consumers, users, and potential users often have a chance to alter or improve upon the technology through pushing back against designers' assumptions, before the "black box" of the technology is closed.[19] It also offers a small and sometimes unlikely–but not impossible–window for rejection of the new technology, in whole or in part.

# Deflating the generative AI bubble

While talk of a bubble has babbled below the surface while the money faucet continues,[20] we observe a recent inflection point, with interlocutors beginning to sound the alarm that AI is overvalued. The perception that AI is a bubble, rather than a goldrush, is making its way into wider discourse with increasing frequency and strength, from actors who are well-positioned and well-regarded in industry as well as politics and culture. The more industry bosses protest that it's not a bubble,[21] the more people have begun to look twice.

Users and artists slammed Adobe for ambiguous statements about using their customer's creative work to train generative AI, forcing the company to later clarify that it would only do so in specific circumstances. At the same time, the explicit promise of not using customer data for AI training has started to become a selling point for others, with a rival positioning their product as "not a trick to access your media for AI training."[22] For other tech companies, a "100% LLM-Free" product that "never present[] chatbot[s] that act human or imitate human experts" is presented as a key feature, spotlighted rather than merely detailed in the terms of service.[23] Lest you think it is only niche companies where such stories emerge, Amazon and Google have also attempted to lower business expectations for Generative AI, because it is expensive, has major accuracy issues, and is as yet an uncertain value proposition.[24] Nonetheless, they have done so in ways that attempt to preserve the hype surrounding AI, which will likely continue to be profitable for their cloud businesses.

---

[15] Halcyon Lawrence explores this dynamic with speech recognition technologies, that for much of their existence were unable to recognize the accents of the majority of English speakers on the planet. See Lawrence, "Siri Disciplines."
[16] Cheng, "$24 Million Iced Tea Company Says It's Pivoting to the Blockchain, and Its Stock Jumps 200%."
[17] Olinga, "Mark Zuckerberg Quietly Buries the Metaverse."
[18] For Google's repeated attempts, see Axon, "RIP (Again): Google Glass Will No Longer Be Sold"; for Apple's recent attempt, see Barr, "Apple Vision Pro U.S. Sales Are All But Dead, Market Analysts Say."
[19] For more on this process see, for example, Kline and Pinch, "Users as Agents of Technological Change."
[20] Celarier, "Money Is Pouring Into AI. Skeptics Say It's a 'Grift Shift.'"
[21] Bratton and Nguyen, "The AI Craze Is No Dot-Com Bubble. Here's Why."
[22] Gray, "Blackmagic Taunts Adobe Following Terms of Use Controversy."
[23] See, for example: "Inqwire."

It is not just technology companies that are asking questions of something they initially presented as inevitable. Recently, venture capital firm Sequoia Capital said "The AI bubble is reaching a tipping point"[25] after not having seen a satisfactory answer to the "Where is all the revenue?" question they asked late last year.[26] Perhaps clearest of all, Goldman Sachs recently published a report titled "Gen AI: too much spend, too little benefit?"[27] in which their global head of equity research stated plainly: "AI technology is exceptionally expensive, and to justify those costs, the technology must be able to solve complex problems, which it isn't designed to do." However, the report tellingly states that even if AI doesn't "deliver on its promise," it may still generate investor returns, as "bubbles take a long time to burst." In short, financial experts are pointing out that capital expenditures on things like graphics cards or cloud compute have not been met by commensurate revenue, nor does there seem to be a clear pathway to remedy this, but some still hold out hope that the period of profit taking has not ended and will not end too soon. This turn in the markets is a recognizable stage in which a product and the set of companies promoting it do not suffer swift devaluation, but begin to lose the ability to hold the top spots on the NASDAQ and other major exchanges.

Why is this happening? Technically, LLMs continue to produce erroneous but confident text ("hallucinate") because they are inherently probabilistic machines, and there are no clear fixes because this is a fundamental feature of how the technology is designed and works.[28] In many cases, LLMs fail to automate the labor that CEOs confidently claimed could be automated, and instead often *decrease* employee productivity.[29] Economically, interest rates have risen, so "easy money" is no longer available[30] to fund boosters' loftiest and horrifically expensive generative AI dreams. Meanwhile, federal regulators have intensified their scrutiny on this niche in the tech economy even as they continue to struggle to reign in social media platforms. FTC chair Lina Khan has said, "There is no AI exemption to the laws on the books," encouraging regulators to fight against AI hype by applying standard regulatory tools and interventions in this new context.[31] Legally, after misappropriating or even allegedly stealing much of their training data in early phases of generative AI development, AI companies are now facing lawsuits, and having to pay more money for their most important input as a result.[32] Public discourse has seemed to catch up to this too: as is common in technology hype cycles, we were promised that AI would automate tedious tasks freeing people for more fulfilling work, even though this seemed counter to reality. Increasingly people who have used these technologies now recognize they are built to, in the worlds of one user, "do my art and writing so that I can do my laundry and dishes," rather than the inverse.[33]

---

[24] Gardizy and Holmes, "Amazon, Google Quietly Tamp Down Generative AI Expectations."
[25] Cahn, "AI's $600B Question."
[26] Cahn, "AI's $200B Question."
[27] Nathan, Grimberg, and Rhodes, "Gen AI: Too Much Spend, Too Little Benefit?"
[28] Leffer, "Hallucinations Are Baked into AI Chatbots."
[29] Robinson, "77% Of Employees Report AI Has Increased Workloads And Hampered Productivity, Study Finds."
[30] Karma, "The Era of Easy Money Is Over. That's a Good Thing."
[31] Khan, "Statement of Chair Lina M. Khan Regarding the Joint Interagency Statement on AI."
[32] O'Donnell, "Training AI Music Models Is about to Get Very Expensive."

# Hype's harmful effects are not easily reversed

While critical commentators on any technology bubble may feel vindicated by seeing it pop, and seeing the "rest of the world"–notably the stock markets–catch up with their gimlet-eyed early critiques, critical technology scholars and the journalists and workers who have been questioning the AI hype also know that the deflation, or even popping of the bubble does not equate to harm undone. Hype has material and often harmful effects in the real world. The ephemerality of these technologies is grounded in real world resources, bodies, and lives in ways that are very similar to the arbitrage methods and resource voracity of industries and corporations at the destructive height of the industrial age. Decades of regulation were required to roll back the environmental and public health harms of technologies we no longer use–from short-lived ones like radium to longer-lived ones like leaded gasoline.[34] Even ephemeral phenomena can have long-lasting negative effects.

The US has slowed plans to retire polluting coal power plants by 40% as a direct result of the hype surrounding AI, with politicians and industry lobbyists justifying this by citing the need for power to win the "AI war".[35] After initially planning to be carbon negative by 2030,[36] Microsoft seemed to drop this after its 2023 emissions were 30% higher than 2020.[37] Brad Smith, its president, said that this "moonshot" goal was made before the "explosion in artificial intelligence", and now "the moon is five times as far away," with AI as the forcing factor. After firing employees[38] for raising concern about Generative AI's high and disparate environmental cost,[39] Google has also seen its emissions increase due to AI, and no longer claims to be carbon-neutral while pushing its net-zero emissions goal date farther into the future.[40] This carbon can't be put back in the ground, or unburned, and the breathless discourse surrounding AI has helped ratchet up the existing climate emergency, providing justification for companies to renege on their already-imperiled environmental promises.[41]

The discourse surrounding AI will also have lasting effects on labor. Some workers will see the scope of their work reduced, while others will see their wages and salaries stagnate or get pushed downward owing to the threat, however empty, that they might be replaced with poor facsimiles of themselves. Creative industries are especially at risk: as illustrator Molly Crabapple states, while demand for a human-created "high" art may remain, generative AI will harm many working illustrators, as editors opt for Generative AI's fast and low cost illustrations over paying artists for original creative output.[42] While artists mobilize with technical and regulatory countermeasures,[43] this burden distracts from their artistic pursuits. Unions such as SAG-AFTRA have won meager protections against AI,[44] and while this hot-button issue perhaps raised the profile of their strike, arguing about generative AI's future role required time and leverage that could have been spent bargaining for wins on other contract points. Even if Generative AI doesn't live up to the hype, its effect on how we value creative work may be hard to shake, with creative workers having to fight to reclaim every inch lost during the Generative AI bubble.

---

[33] Maciejewska, "I Want AI to Do My Laundry and Dishes so That I Can Do Art and Writing…"
[34] Clark, *Radium Girls, Women and Industrial Health Reform*; Nader, *Unsafe at Any Speed*.
[35] Chu, "US Slows Plans to Retire Coal-Fired Plants as Power Demand from AI Surges."
[36] Smith, "Microsoft Will Be Carbon Negative by 2030."
[37] Rathi, Akshat, Bass, Dina, and Rao, Mythili, "A Big Bet on AI Is Putting Microsoft's Climate Targets at Risk."
[38] Simonite, "What Really Happened When Google Ousted Timnit Gebru."
[39] Bender and Gebru et al., "On the Dangers of Stochastic Parrots."
[40] Rathi, "Google Is No Longer Claiming to Be Carbon Neutral."
[41] Kneese and Young, "Carbon Emissions in the Tailpipe of Generative AI."

Lastly, generative AI will also have long-term effects on our information commons. The ingestion of massive amounts of user-generated data, text, and artwork–often in ways that appear to violate copyright and fair use–has pushed us further along the road to the enclosure of the information commons by corporations.[45] Google's AI search snippets tool authoritatively suggested putting glue in pizza, and said that experts recommend eating at least one small rock per day.[46] While these may seem obvious enough to be harmless, most AI generated misinformation is not so obvious. The increasing prevalence of AI generated nonsense on the internet will make it harder to find trusted information, allow misinformation to propagate, and erode trust in places we used to count on for reliable information.

A key question remains for which we may never have a satisfactory answer: what if the hype was always meant to fail? What if the point was to hype things up, get in, make a profit, and entrench infrastructure dependencies before critique, or reality, had a chance to catch up?[47] Path dependency is well understood by historians of technology and those seeking to profit from AI alike: today's hype will have lasting effects that constrain tomorrow's possibilities. Using the AI hype to shift more of our infrastructure to the cloud increases dependency on cloud companies, and this will be hard to undo even as inflated promises for AI are dashed.

Inventors, technologists, corporations, boosters, and investors regularly seek to create inevitability, in part by encouraging a discourse of inexorable technological "progress" tied to their latest investment vehicle. They do this in part by referencing past technologies, which seem natural and necessary from the vantage point of the present, claiming their development (tautologically) as inevitable. Yet the efforts to make AI indispensable on a large scale, culturally, technologically, and economically, have not lived up to their promises, and in a sense this is not surprising. Indeed, Generative AI does not so much represent the wave of the future as it does the ebb and flow of waves past.

---

[42] Crabapple and Marx, "Why AI Is a Threat to Artists, with Molly Crabapple."
[43] Jiang et al., "AI Art and Its Impact on Artists."
[44] Frawley, "Unpacking SAG-AFTRA's New AI Regulations."
[45] See Noble, *Algorithms of Oppression*, for a fuller discussion of the reconfiguration of the information landscape in the US (and global) online ecosystem to be firmly under the control of for-profit companies making billions, mostly off advertising.
[46] Kelly, "Google's AI Recommends Glue on Pizza: What Caused These Viral Blunders?"
[47] Some financial self regulatory authorities have even added warnings about AI pump-and-dump schemes, see for example: Financial Industry Regulatory Authority, "Avoid Fraud: Artificial Intelligence (AI) and Investment Fraud."

# Acknowledgements

We are grateful to Ali Alkhatib, Sireesh Gururaja, and Alex Hanna for their insightful comments on earlier drafts.